\documentstyle[preprint,epsfig,aps]{revtex}
 \tighten
\def\bea{\begin{eqnarray}}
\def\beann{\begin{eqnarray*}}
\def\beq{\begin{equation}}
\def\eea{\end{eqnarray}}
\def\eeann{\end{eqnarray*}}
\def\eeq{\end{equation}}

\begin{document}
\draft
\title{Influence of a $Z^+$(1540) resonance on $K^+N$ scattering}
\author{
J. Haidenbauer$^{1}$, and G. Krein$^{2}$ \\
{\small $^1$ Forschungszentrum J\"ulich, Institut f\"ur Kernphysik, 
D-52425 J\"ulich, Germany} \\
{\small $^2$ Instituto de F\'{\i}sica Te\'{o}rica, Universidade Estadual 
Paulista } \\
{\small Rua Pamplona, 145 - 01405-900 S\~{a}o Paulo, SP, Brazil} 
}
\maketitle
\begin{abstract}
The impact of a ($I=0$, $J^P={\frac{1}{2}}^+$) $Z^+$(1540) resonance with
a width of 5 MeV or more on the $K^+N$ (I=0) elastic cross section and
on the $P_{01}$ phase shift is examined within the $KN$ meson-exchange
model of the J\"ulich group. It is shown that the rather 
strong enhancement of the cross section caused by the presence of a 
$Z^+$ with the above properties is not compatible with the 
existing empirical information on $KN$ scattering. 
Only a much narrower $Z^+$ state could be reconciled with
the existing data -- or, alternatively, the $Z^+$ state must 
lie at an energy much closer to the $KN$ threshold. 
\end{abstract}
\vspace{2.5cm}
\noindent{PACS NUMBERS: 13.75.Jz, 12.39.Pn, 14.20.Jn, 21.30.-x, 12.40.-y}

\vspace{1.0cm}

\newpage 
%


Recently 
the LEPS collaboration at Spring-8 presented evidence for the existence 
of a narrow baryon resonance with strangeness $S=+1$ \cite{spring}. 
In the following four other collaborations from different laboratories announced 
the observation of a similar structure in their experiments 
\cite{itep,jlab,elsa,itep2}. 
The observed structure was immediately brought into connection 
with an exotic pentaquark state called 
$Z^+$ whose existence had been proposed since long time in the context 
of different quark models~\footnote{We follow the historical 
nomenclature adopted in the particle data tables. More recently the 
resonance is being called $\Theta^+$.}. 
Specifically, the resonance parameters with a peak position  
around $1540$~MeV and a width around $20$~MeV,
extracted from these experiments, lie convincingly close to 
a theoretical prediction based on the chiral quark-soliton model of
Diakonov et al.~\cite{dpp}, who had proposed the existence of a $Z^+$ state 
with a mass around $1530$~MeV and a width of around $15$ MeV. Due to its
quantum numbers, $S=+1$, $I=0$, and $J^P={\frac{1}{2}}^+$, their $Z^+$ state
can only decay (hadronically) into the $K^+n$ or $K^0p$ channels.

First cautious words about this interpretation were, however, raised by
S. Nussinov \cite{nussinov} soon after the experimental result \cite{spring}
was published. He pointed out that the existence
of such a $Z^+$ state at around 1540 MeV should also be seen in the
available $K^+d$ scattering data. Though some ``intriguing
fluctuations'' exist in the total $K^+d$ cross section in the energy range
which corresponds to $KN$ cms energies of 1500-1600 MeV \cite{nussinov} 
Nussinov's conclusion
was that the lack of a prominent $Z^+$ signature in $K^+d$ collisions
restricts the width of the $Z^+$ to be smaller than 6 MeV. 
Similar but even more restrictive conclusions were drawn not long afterwards
by Arndt and collaborators~\cite{Arndt}. These authors reexamined 
the available $K^+N$ scattering data basis with the aim of exploring the 
possibility of accommodating a $Z^+$-like resonance structure in their
partial wave analysis. In an earlier
analysis of the same data by the VPI group~\cite{vpi} this resonance has not
been explicitly considered. 
The work of Ref.~\cite{Arndt} confirmed that the existing $K^+N$ data 
excludes $Z^+$ widths beyond the few-MeV level. Indeed their results 
even suggest that a $Z^+$ around 1540 MeV should have a width of 
$\Gamma$ = 1 MeV or even less in order to be compatible with the
$K^+N$ and $K^+d$ data basis.  

In the present note we use the J\"ulich meson-exchange model for the $KN$ 
interaction to investigate the effect of including in the model a $Z^+$-like 
resonance structure on the description of the experimental data. 
Within a realistic potential model the open
parameters are fixed by a simultaneous fit to all
$KN$ partial waves and therefore the contributions to the $P_{01}$
channel (we use the standard spectral notation $L_{I \ 2J}$),
which provide the background for the $Z^+$(1540) resonance,
are strongly constrained by the empirical information in the other
partial waves and that means also from the other isospin channel. 
Furthermore, the use of a model allows one to produce a resonance structure 
from a bare pole interaction by dressing the bare baryon-meson vertex, with a 
width generated from self-energy loops, i.e. the non-pole and the pole part
of the reaction amplitude can be treated consistently. 
 
A detailed description of the J\"ulich $KN$ 
model can be found in Refs.~\cite{Juel1,Juel2}. The model was constructed along
the lines of the (full) Bonn $NN$ model \cite{MHE} and its extension 
to the hyperon-nucleon ($YN$) system \cite{Holz}. 
Specifically, this means that one
has used the same scheme (time-ordered perturbation theory), the same 
type of processes, and vertex parameters (coupling constants, cut-off
masses of the vertex form-factors) fixed already by the study of 
these other reactions. 

The diagrams considered for the $KN$ interaction are shown in
Fig.~\ref{Juel_mod}. Obviously the J\"ulich model contains not only 
single-meson (and baryon) exchanges (Fig.~\ref{Juel_mod}a), but also 
higher-order box diagrams involving $NK^*$, $\Delta K$ and $\Delta K^*$ 
intermediate states (Fig.~\ref{Juel_mod}b). 
Based on these diagrams a $KN$ potential $V$ is
derived, and the corresponding reaction amplitude
$T$ is then obtained by solving a Lippmann-Schwinger type 
equation defined by time-ordered perturbation theory:
\begin{equation}
T = V + V G_0 T \ .
\label{LSE}
\end{equation}
From the reaction amplitude $T$ phase shifts and observables (cross sections,
polarizations) can be obtained in the usual way.

In the present investigation we use the $KN$ model I described 
in Ref. \cite{Juel2}. (Note that we have performed also exploratory
calculations with the other models in Refs. \cite{Juel2,HHK} 
and we obtained essentially the same results.) 
Results for phase shifts and also for cross sections and polarizations
can be found, e.g., in Ref. \cite{Juel2}.  
Evidently this model yields a good overall reproduction of all
presently available empirical information on $KN$ scattering. 
Specifically, it describes the data up to beam momenta of $p_{lab}
\ \approx$ 1 GeV/c, i.e. well beyond the region of the observed 
$Z^+$(1540) resonance structure which corresponds to the momentum $p_{lab}$ = 
0.44 GeV/c. Thus, this model provides a solid basis for studying the
influence of the $Z^+$(1540) resonance on the $KN$ observables. 
As already emphasized above 
the parameters of the model are fixed by a simultaneous fit to all
$KN$ partial waves and therefore the contributions to the $P_{01}$
channel, where the $Z^+$ pentaquark state is supposed to occur \cite{dpp},
are constrained by the empirical information in the other
partial waves. 
 
The $Z^+$(1540) resonance is included in the model by 
adding a pole diagram, as depicted in Fig. \ref{Juel_mod}c, 
with a bare mass $M^{(0)}_{Z^+}$ and a bare coupling constant
$g^{(0)}_{KNZ^+}$ to the other diagrams that contribute to $V$. When
this interaction is then iterated in the Lippmann-Schwinger equation (\ref{LSE}) 
the ${KNZ^+}$ vertex gets dressed by the non-pole part of the interaction 
and the $Z^+$ acquires a width and also its physical mass via self-energy
loops. For the present investigation we prepared two different models, 
one with a width of 20 MeV, as found in the experiment
\cite{jlab}, and one
with a width of just 5 MeV, that was given in Ref. \cite{Pol00} as 
the most favorable width of the chiral quark-solition model,
and which corresponds roughly to the upper limit given in the paper
by Nussinov \cite{nussinov}. The width of the $Z^+$ and also the 
resonance mass are calculated from a speed plot, 
but we must say that for such a narrow structure the 
resonance position basically coincides with the energy where the phase
passes through 90 degrees. Note that the bare mass and bare coupling constant
are free parameters that 
are used to adjust the desired physical mass and width of the $Z^+$. The cutoff
mass occurring in the vertex form factor, cf. Eq. (2.23) of Ref. \cite{Mueller},
was fixed to 2 GeV. 
 
The elastic cross sections (for the isospin channels $I=0,1$) predicted by
the two models with a $Z^+$ are shown in Fig. \ref{XS} together with the
results of the original J\"ulich model I and the available experimental 
information \cite{Bow70,Coo70,Car73,Bow73}. 
The $I=1$ channel is shown here only to demonstrate the quality
of the J\"ulich model. The $Z^+$ is, of course, assumed to be a $I=0$
resonance and therefore it does not change the results in the $I=1$ channel. 
 
The J\"ulich model provides also a decent description of the data in 
the $I=0$ channel. Its prediction might lie slightly too low at higher energies,
however, one has to take into account that the data scatter also somewhat. 
In any case, it is obvious that the deviation of the J\"ulich model and
also the variations between the different data sets are by no means
comparable to the impact of the $Z^+$ on the $KN$ ($I=0$) cross section. 
It is also clear that the $Z^+$ as predicted by \cite{dpp} and as 
supposedly seen in the experiments \cite{spring,itep,jlab,elsa,itep2} lies
well within the energy range covered by $KN$ data. Indeed there are
even data points from two independent experiments \cite{Bow70,Car73}. 

There is no way to reconcile the present $KN$ ($I=0$) cross section 
data with the existence of a $Z^+$(1540) with a width of 5 MeV or more. 
In view of the curves shown in Fig. \ref{XS} it is clear why 
Arndt and collaborators saw such a strong increase of the $\chi^2$
in their partial wave analysis once the $Z^+$ (with $\Gamma$ = 5 MeV
or more) was included \cite{Arndt}. One of their conclusions was
that the $Z^+$ could have a width of order 1 MeV or less. We did not
consider such a small width within our model. However, it is clear
that reducing the width significantly would eventually lead to 
results that coincide with the ones of the J\"ulich model -- besides
an isolated narrow peak somewhere. Since there are no data below
$p_{lab}$ = 0.336 GeV/c there is indeed also room for the $Z^+$
at energies much closer to the $KN$ threshold. However, then one
would need to find a dynamical explanation why the structure seen
in the experiments \cite{spring,itep,jlab,elsa,itep2} appears at a
significantly higher $KN$ invariant mass there -- provided, of
course, that it has something to do with the pentaquark  
state predicted in Ref. \cite{dpp}. 

Results for the $P_{01}$ phase shift are shown in Fig. \ref{phase}. 
Note that the phases for the models with the $Z^+$ resonance pass
through 90 degrees around $p_{lab}$ = 0.44 GeV/c and then continue
to rise beyond 180 degrees. Therefore we show them here modulo $\pi$
so that they fit on the same graph and approach the J\"ulich model
and the results of the phase shifts analyses again at higher energies. 
Also here it is clear that the existence of a $Z^+$ with a width of
5 MeV or more would lead to a tremendous change. 

In summary, we have demonstrated the impact of a $Z^+$(1540) with
a width of 5 MeV or more on the $KN$ (I=0) elastic cross section and
on the $P_{01}$ phase shift. Even though the $KN$ data in the relevant
energy range show sizeable uncertainties it is evident that the rather 
strong enhancement of the cross section caused by the presence of a 
$Z^+$(1540) would be in clear contradiction to the experiments. 
Only a much narrower $Z^+$(1540) state could be reconciled with
the existing empirical information on $KN$ scattering 
\cite{nussinov,Arndt} -- or the predicted pentaquark state must 
occur at an energy much closer to the $KN$ threshold. In any case
it would be desirable to re-measure $KN$ scattering around the
energy of the suspected $Z^+$(1540) resonance using present-days
much more advanced accelerators and detector systems. 

\acknowledgements{Work partially financed by CNPq and FAPESP.}

\newpage

\begin{figure}
\caption{Meson-exchange contributions to the $KN$ interaction. 
Diagrams (a) and (b) define the original J\"ulich model I 
\protect\cite{Juel1,Juel2} that we use in the present investigations. 
Diagram (c) represents the considered $Z^+$(1540) contribution.
}
\label{Juel_mod}
\end{figure}

\begin{figure}
\caption{$KN$ elastic cross section in the isospin channels I=0,1. 
The solid line is the 
result of the original J\"ulich model I from Ref.~\protect\cite{Juel2}.
The dashed (dash-dotted) line shows results where a $Z^+$ resonance with a 
dynamically generated width of 5 (20) MeV is included. 
Experimental data are taken from Ref. \protect\cite{Bow70} (filled circles),
Ref. \protect\cite{Coo70} (open squares),
Ref. \protect\cite{Car73} (open circles),
and Ref. \protect\cite{Bow73} (crosses).}
\label{XS}
\end{figure}

\begin{figure}
\caption{
$KN$ phase shifts in the $P_{01}$ partial wave. 
The solid line is the 
result of the original J\"ulich model I from Ref.~\protect\cite{Juel2}.
The dashed (dash-dotted) line shows results where a $Z^+$ resonance with a 
dynamically generated width of 5 (20) MeV is included. 
Experimental phase shifts are
taken from Ref.~\protect\cite{Exp1} (open circles),
Ref.~\protect\cite{Exp2} (open squares), and
Ref.~\protect\cite{Exp3} (filled circles and pluses)
.}
\label{phase} 
\end{figure}

\newpage 

\begin{figure}
\centerline{\epsfxsize=12.0cm\epsfysize=15.0cm\epsfbox{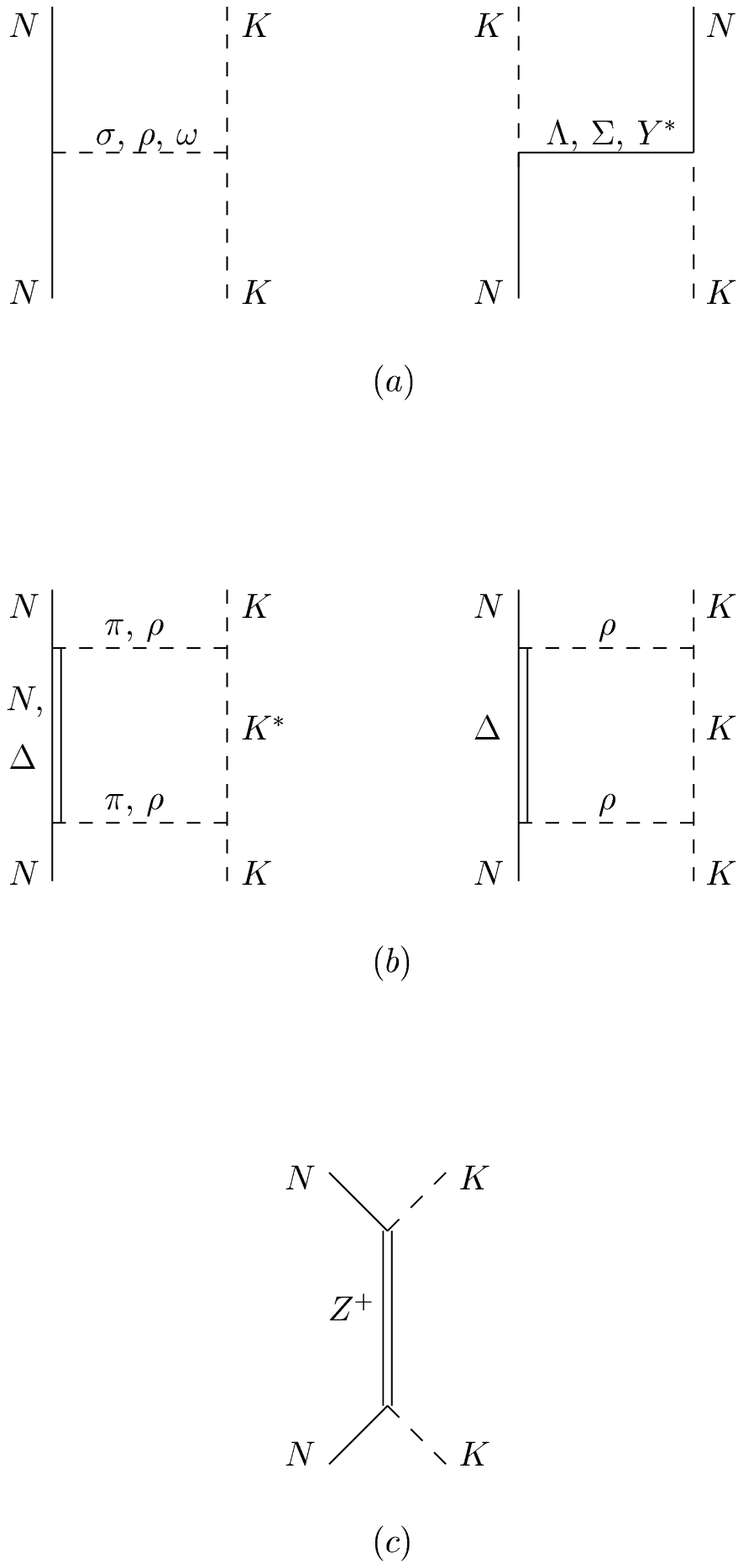} }
\end{figure}
\center{Fig. 1}

\newpage

\vglue 1cm
\begin{center}
\begin{figure}
\epsfig{file=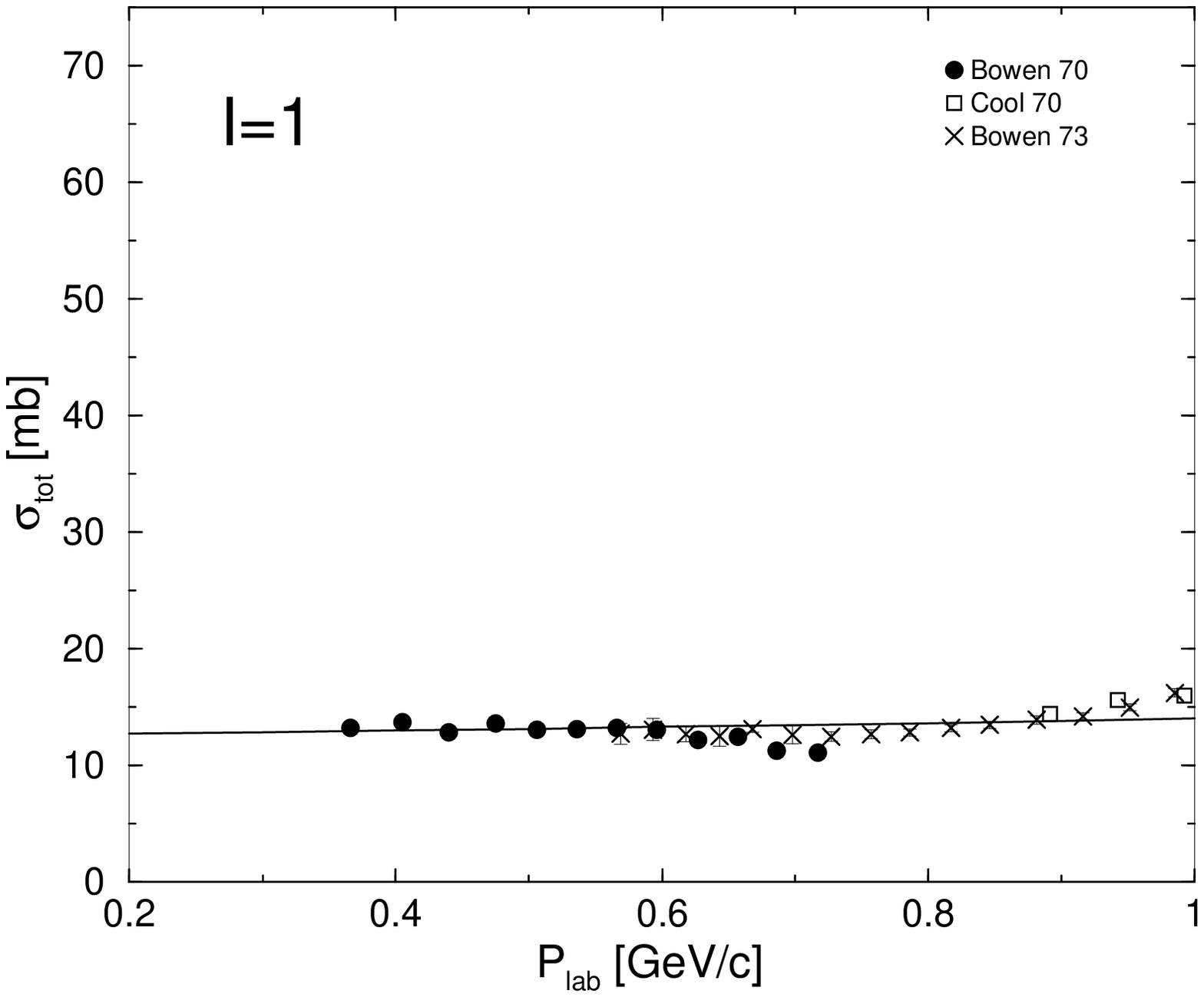, width=11.0cm}
\epsfig{file=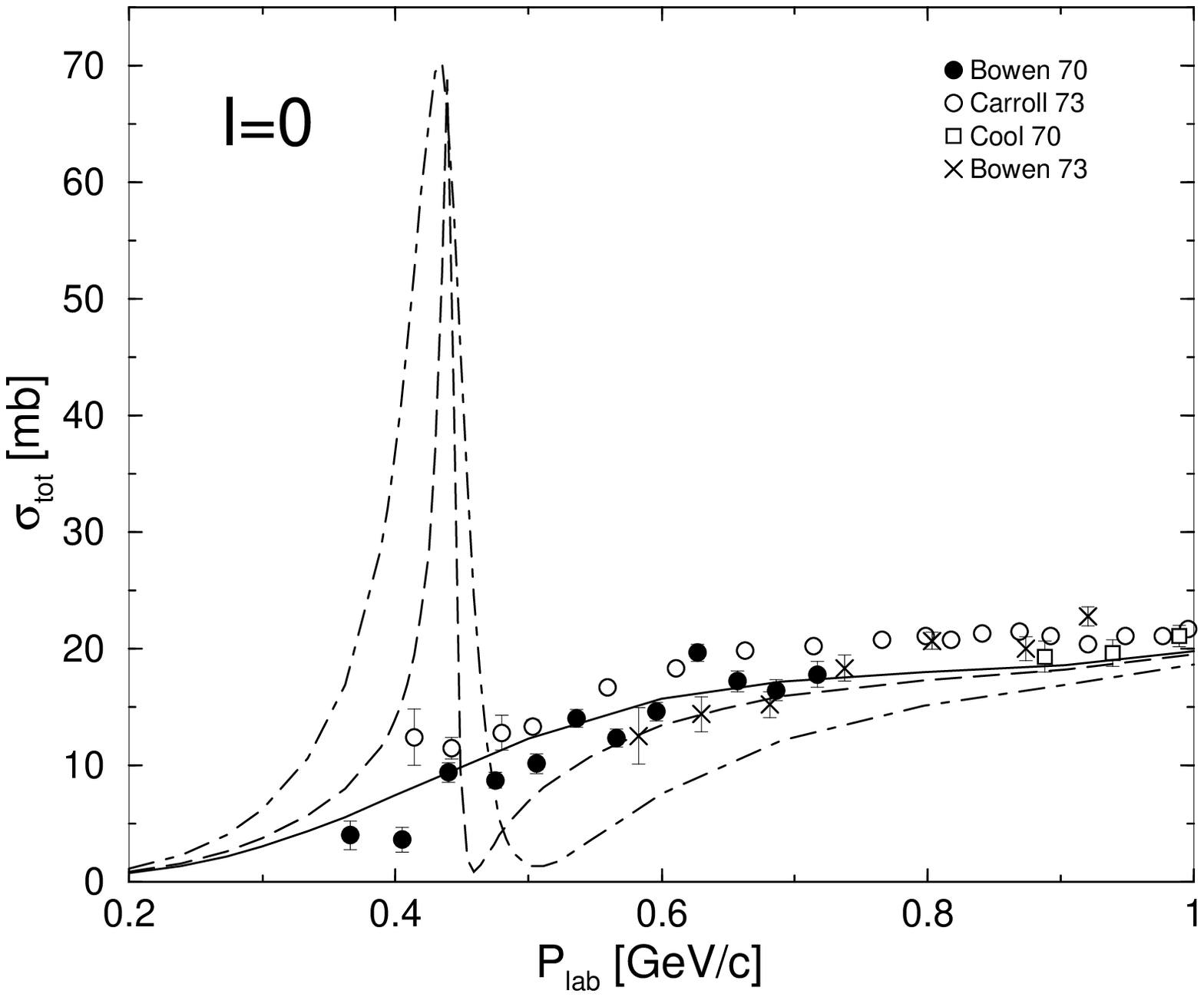, width=11.0cm}
\end{figure}
\end{center}
\center{Fig. 2}

\newpage

\vglue 1cm
\begin{center}
\begin{figure}
\epsfig{file=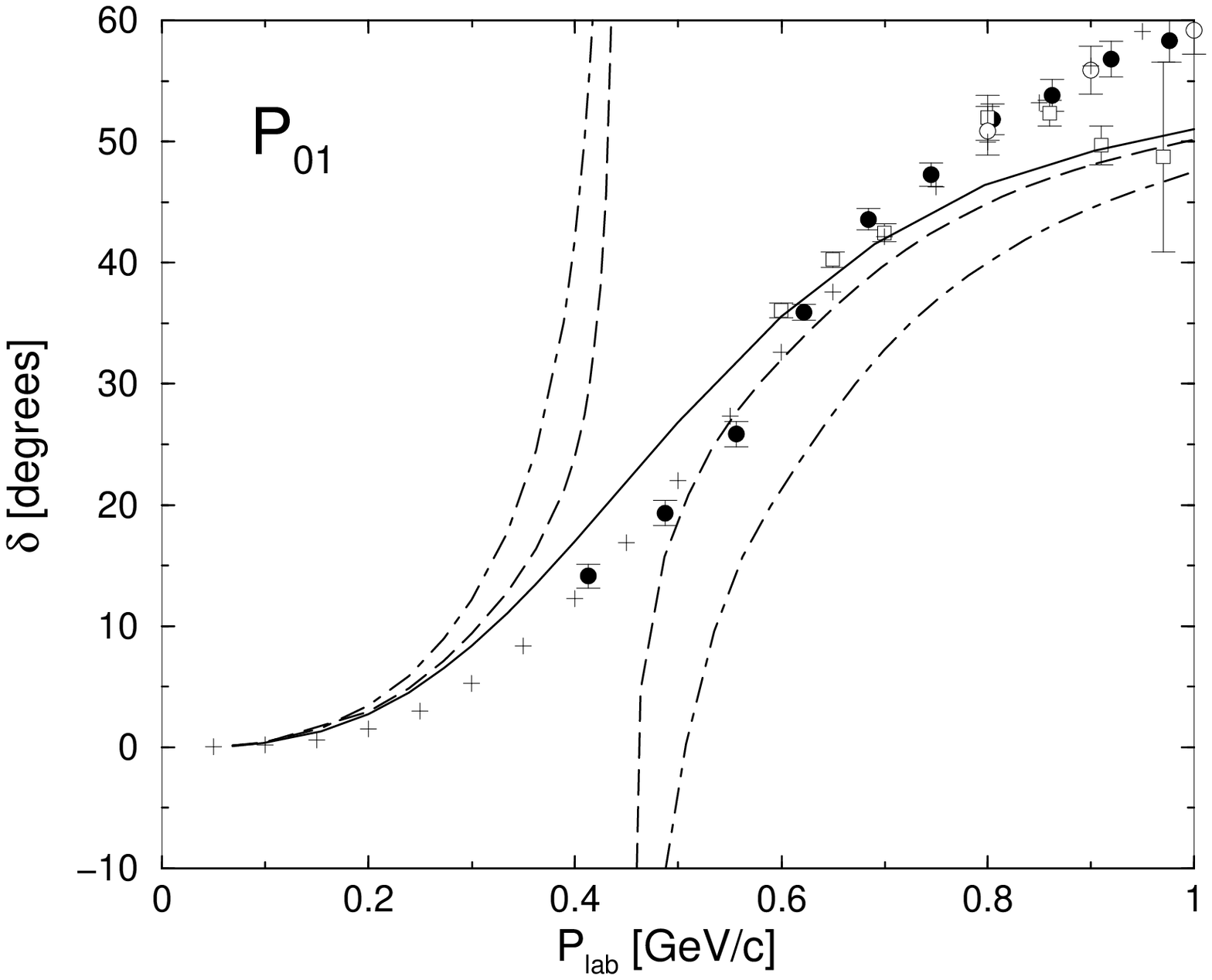, width=11.0cm}
\end{figure}
\end{center}
\center{Fig. 3}

\end{document}